# A Resonant soft x-ray powder diffraction study to determine the orbital ordering in A-site ordered $SmBaMn_2O_6$

M. García-Fernández[1], U. Staub[1], Y. Bodenthin[1], S. M. Lawrence[2], A. M. Mulders[2,3], C. E. Buckley[2], S. Weyeneth[4], E. Pomjakushina[5, 6], and K. Conder[5]

[1]Swiss Light Source, Paul Scherrer Institut, 5232 Villigen PSI, Switzerland

[2]Department of Imaging and Applied Physics, Curtin University of Technology, Perth, WA 6845, Australia

[3]The Bragg Institute, Australian Nuclear Science and Technology Organization, Lucas Heights, NSW 2234, Australia

[4]Physik-Institut der Universität Zürich, Winterthurerstrasse 190, 8057 Zürich, Switzerland

[5]LDM, Paul Scherrer Institut, 5232 Villigen PSI, Switzerland

[6] LNS, PSI & ETH Zürich, 5232 Villigen PSI, Switzerland

Abstract

Soft X-ray resonant powder diffraction has been performed at the Mn $L_{2,3}$ edges of A-site ordered $SmBaMn_2O_6$. The energy and polarization dependence of the (¼ ¼ 0) reflection provide direct evidence for a $(x^2-z^2)/(y^2-z^2)$ type orbital ordering in contrast to the single layer manganite. The temperature dependence of the reflection indicates an orbital reorientation transition at ≈210 K, below which the charge and orbital ordered $MnO_2$ sheets show AAAA type of stacking. The concurring reduction of the ferromagnetic super exchange correlations leads to further charge localization.

Doped manganites have attracted considerable attention as they show a variety of electronic and magnetic ground states ranging from states with colossal magnetic resistance to charge and orbital ordered ground states.[1] The rich variety of different ground states exemplifies the strong interactions between the magnetic and electronic degrees of freedom, which are coupled to the crystal lattice. At half doping, 3-dimensional $R_{1-x}Ca_xMnO_3$ (R=Y or rare earth) perovskites are often antiferromagnetic insulators with charge and orbitally ordered ground states.[2] The $R_{0.5}Ba_{0.5}MnO_3$ ($RBaMn_2O_6$) systems have significant different ground states depending on the randomness of the A-site R/Ba substitution.[3] For a layered ordering of the R and Ba ions, a long range charge ordered (CO) and orbitally ordered (OO) ground state is observed, whereas for the disordered materials, no CO/OO ordering occurs and a magnetic glassy state is observed. The importance of disorder is also displayed by the observation of finite correlations in the orbital ordered state of the Ca doped solid solutions.[4]

The A-site ordered manganite $RBaMn_2O_6$ is an ideal candidate to study the long range orbital and charge order in detail. The crystal structure of the cation-ordered material is of $a_p$ x $a_p$ x $2a_p$ type, with $a_p$ the cubic perovskite unit cell. The temperature for charge and orbital order depend strongly on the size mismatch of the R/Ba ordering. For instance for R=Y, the metal insulator transition at $T_{CO}$ = 500 K is much higher than found for other manganites. The tetragonal crystal structure symmetry is lowered at $T_{CO}$ to C2/m.[5] For R=Sm, the material considered in this work, $T_{CO} \approx 360$ K and an antiferromagnetic transition occurs at $T_N \approx 250$ K, followed by a reorientation of the orbital order at $T_{CO2} \approx 200$K as proposed from electron diffraction.[6, 7] High-resolution neutron powder diffraction[5, 8] found that this transition is accompanied by a structural transition lowering the symmetry to P-1. These studies also propose a ferro-orbital ordering for R=Y in the temperature region $T_{CO} < T < T_{CO2}$.

A direct study of the orbital and charge ordered ground state is achieved with resonant x-ray diffraction. Experiments performed in the vicinity of the Mn *K* edge are mainly sensitive to the Jahn-Teller distortion,[9] while experiments in the vicinity of the Mn *L* edge directly probe the transition metal 3*d* states and are most sensitive to charge, orbital and magnetic order.[10] These soft X-ray resonant diffraction studies have directly observed the OO but so far no agreement on its type, i.e. $x^2-z^2/y^2-z^2$ versus $3x^2-r^2/3y^2-r^2$, has been achieved for the single layer manganite.[11-14] However, these studies showed the significance of Mn-O hybridization and its influence on the spin correlations.[12, 14] In addition, the electronic orbital order and the associated Jahn-Teller distortion demonstrate an individual temperature dependence distinguished via their specific resonant energies.[15]

In this paper we present the first resonant soft X-ray powder diffraction experiments of an orbital order reflection. Small but distinct differences in the observed energy spectra reveal an $x^2-z^2/y^2-z^2$ type of ordering in contrast with $La_{0.5}Sr_{1.5}MnO_4$. The doubling in intensity of the orbital reflection at $T_{CO2}$ gives direct evidence for an orbital reorientation transition leading to AAAA type stacking and which causes (or is caused by) the suppression of ferromagnetic correlations and further localization of charge carriers.

The sample of $SmBaMn_2O_6$ was synthesized by a solid state reaction using $Sm_2O_3$, $BaCO_3$ and $MnO_2$ of a minimum purity of 99.99%. The respective amounts of starting reagents were mixed and calcinated at temperature of 1400$^o$C for 40h in an argon flow, with several intermediate grindings, then pressed in pellets, sintered at 1400$^o$C for 10h in an argon flow and oxidized in an oxygen flow at 700$^o$C for 4h. Phase purity of the synthesized compound was checked by Cu$K_\alpha$ x-ray diffraction. Resonant soft x-ray diffraction experiments were performed on the RESOXS endstation at the SIM beamline of the Swiss Light Source of the Paul Scherrer Institut, Switzerland. A polycrystalline pellet of 10 mm diameter was glued onto a copper sample holder mounted on a He flow cryostat, which achieves temperatures between 10 K and 370K. Experiments were performed using linear horizontal or vertical polarization light leading to $\pi$ or $\sigma$ incident photon polarization in the horizontal scattering geometry, respectively. Two dimensional data sets were collected with a commercial Roper Scientific CCD camera mounted in vacuum.

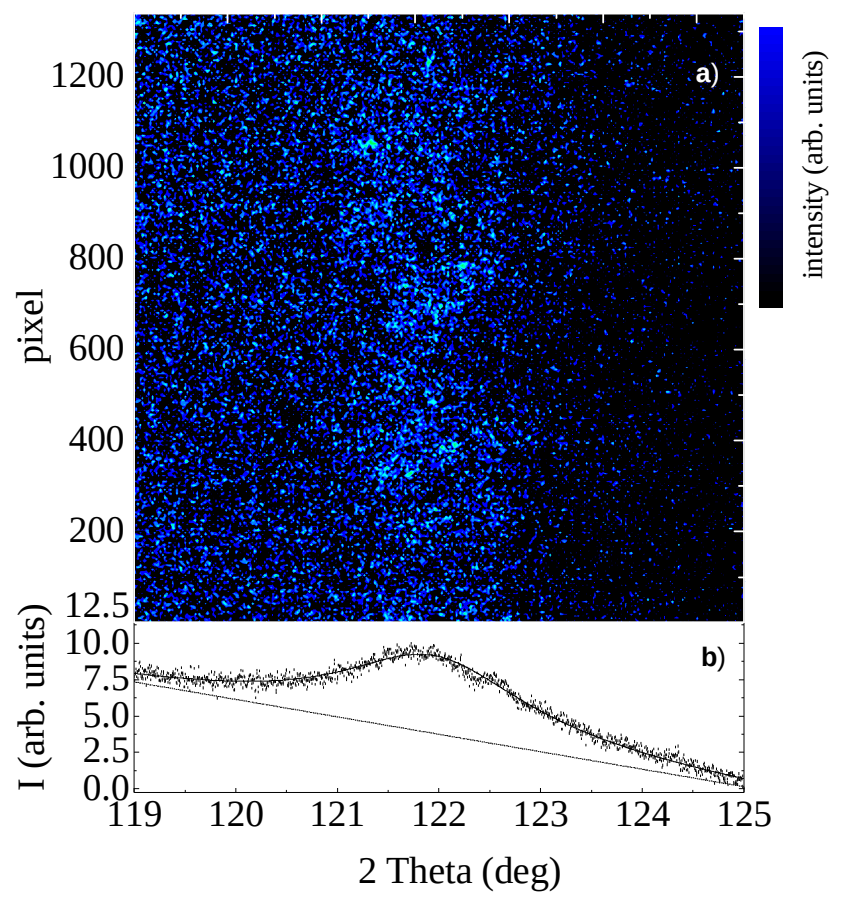

FIG. 1 a) Image of a section of the resonant orbital powder diffraction ring (¼ ¼ 0) taken at the Mn $L_3$ edge (643.25 eV) at 14K with π incident radiation located at 2θ of approximately 120 degree of $SmBaMn_2O_6$. b) Integrated intensity (vertical) of the image as a function of 2θ.

An image of a section of the powder ring of the (¼ ¼ 0) reflection of $SmBaMn_2O_6$ at the Mn $L_3$ edge at $T$=14 K is shown in Figure 1a and a one-dimensional integration of the image is shown in Figure 1b. The integrated intensity is obtained with a fit to a Lorentz function representing the (¼ ¼ 0) reflection and a linear background accounting for the fluorescence (see Figure 1b). The energy dependence of the fluorescence and the integrated intensity of the orbital (¼ ¼ 0)) reflection at the Mn $L_{2,3}$ edges are shown in Figure 2a and compared with those of layered $La_{0.5}Sr_{1.5}MnO_4$. There is a strong similarity between the two; all the main features present in the spectra of the layered manganite are also present in the spectra of $SmBaMn_2O_6$, though the intensity and energy separation are slightly different. Two significant dissimilarities are the much weaker shoulder of the $SmBaMn_2O_6$ to the right of the main feature of the $L_3$ edge (B*) and the missing shoulder to the left of feature E (E*) at the $L_2$ edge. These two features are also distinct when comparing the theoretical calculations for the $x^2$-$z^2$/$y^2$-$z^2$ and $3x^2$-$r^2$/$3y^2$-$r^2$ types of OO as presented by two different groups with different approximations.[13, 14] In both cases, the shoulder B* is much weaker (or at higher energy) and the feature E* is absent for the $x^2$-$z^2$/$y^2$-$z^2$ type of OO (see Figure 2b inset). This indicates that the OO of the $e_g$ electrons is of $x^2$-$z^2$/$y^2$-$z^2$ type in the layered $SmBaMn_2O_6$ while a $3x^2$-$r^2$/$3y^2$-$r^2$ type of orbital ordering is present in the single layer manganite.

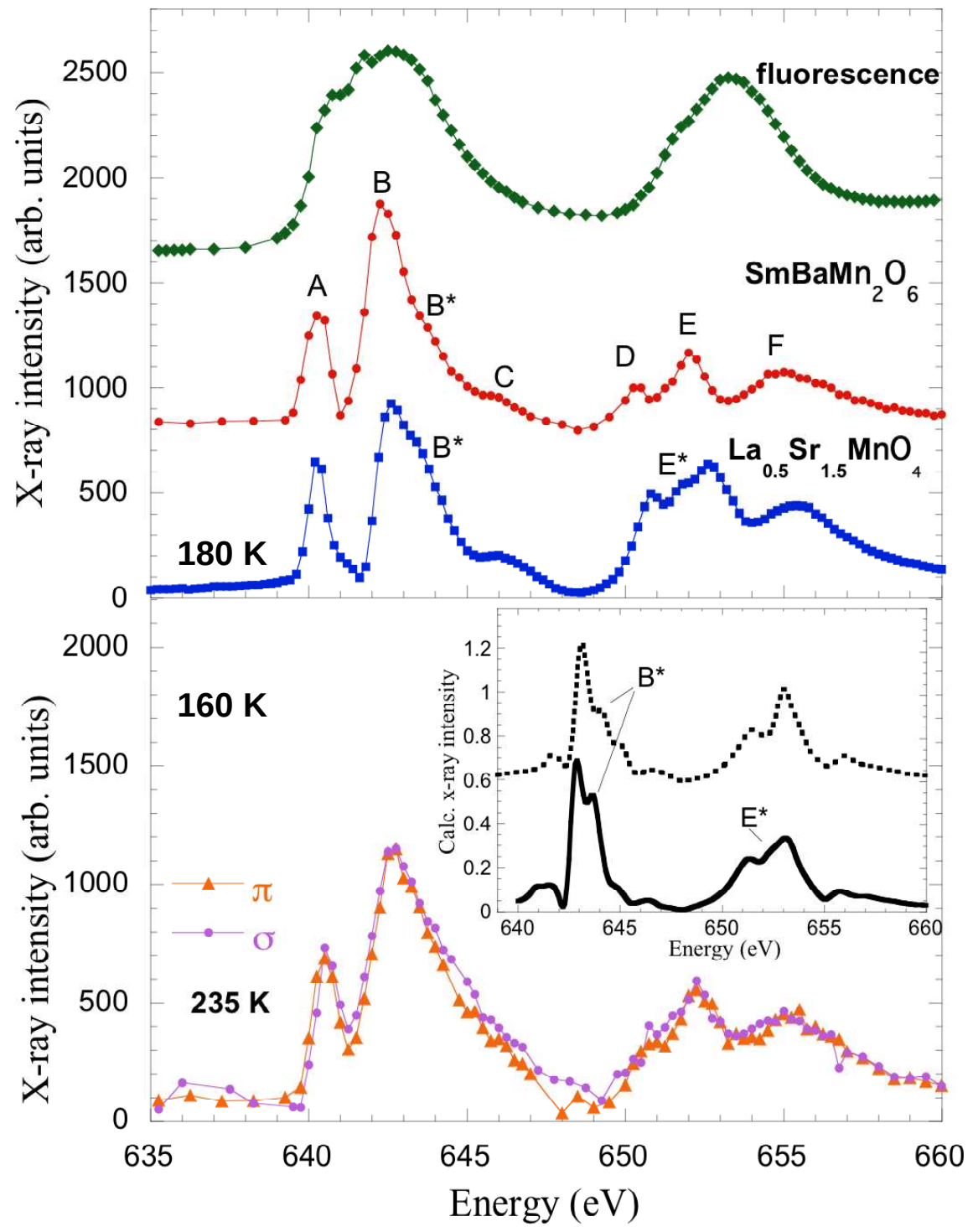

FIG. 2 (color online) a) Energy dependence of the fluorescence and integrated intensity of the (¼ ¼ 0) reflection of polycrystalline $SmBaMn_2O_6$ with π incident radiation at 120K compared with that of single crystal $La_{0.5}Sr_{1.5}MnO_4$ taken at 160K (the intensity is scaled and shifted for easy comparison).[15] b) Energy dependence of the (¼ ¼ 0) reflection recorded with π and σ incident radiation at 235K. Inset: Calculated Energy dependence of the (¼ ¼ 0) reflection for $x^2$-$y^2$ and $3z^2$-$r^2$ type OO taken from Ref. 14.

The spectra taken with σ and π polarization are equal within experimental accuracy (Figure 2b), indicating that the scattered radiation is rotated (σ–π or π–σ), because signals in the unrotated channels (σ–σ and π–π) are different from each other. For $La_{0.5}Sr_{1.5}MnO_4$ this has been proven directly using polarization analysis of the scattered radiation,[15] which is not feasible in this study due to the weak intensity of the powder reflection. Though the polarization behavior of the (1/4, 1/4, 0 ) reflection contains inherent information on the orbital ordering, e.g. the site symmetry of the Mn ions, it unfortunately does not discriminate between the two orbital order models.

The temperature dependence of the orbital (¼ ¼ 0) reflection taken at the maximum of the $L_3$ edge (642.25 eV) is shown in Figure 3a. The orbital reflection appears at $T_{CO} \approx 355$ K and its intensity is constant between 330 K and 220 K. Below $T_{CO2} \approx 210$ K a sharp increase in intensity is observed and the intensity at 180 K is twice that of the intensity above $T_{CO2}$. The intensity increases further below 180 K but with a smaller gradient. The integrated intensity reflects the orbital order parameter and is compared with the magnetization and the resistivity in Figure 3b. The magnetization and resistivity of our sample are similar to those reported in ref [6]. Below $T_{CO}$ a significant increase in the resistance and decrease in the magnetization is observed. Below approximately 240 K the field cooled (FC) and zero field-cooled (ZFC) magnetizations strongly increase due to antiferromagnetic ordering of the Mn spins with a small spin canting resulting in a weak ferromagnetic component. The weak ferromagnetic component is clearly visible in the hysteresis behavior of the FC scans between 100 K and 240 K. At $T_{CO2} \approx 210$K the FC magnetization has a maximum and a change in slope of the logarithmic plotted resistance is evident. We conclude that those features as well as the increase in (¼ ¼ 0) intensity are due to a change in the orbital stacking pattern along the c-direction [6, 7].

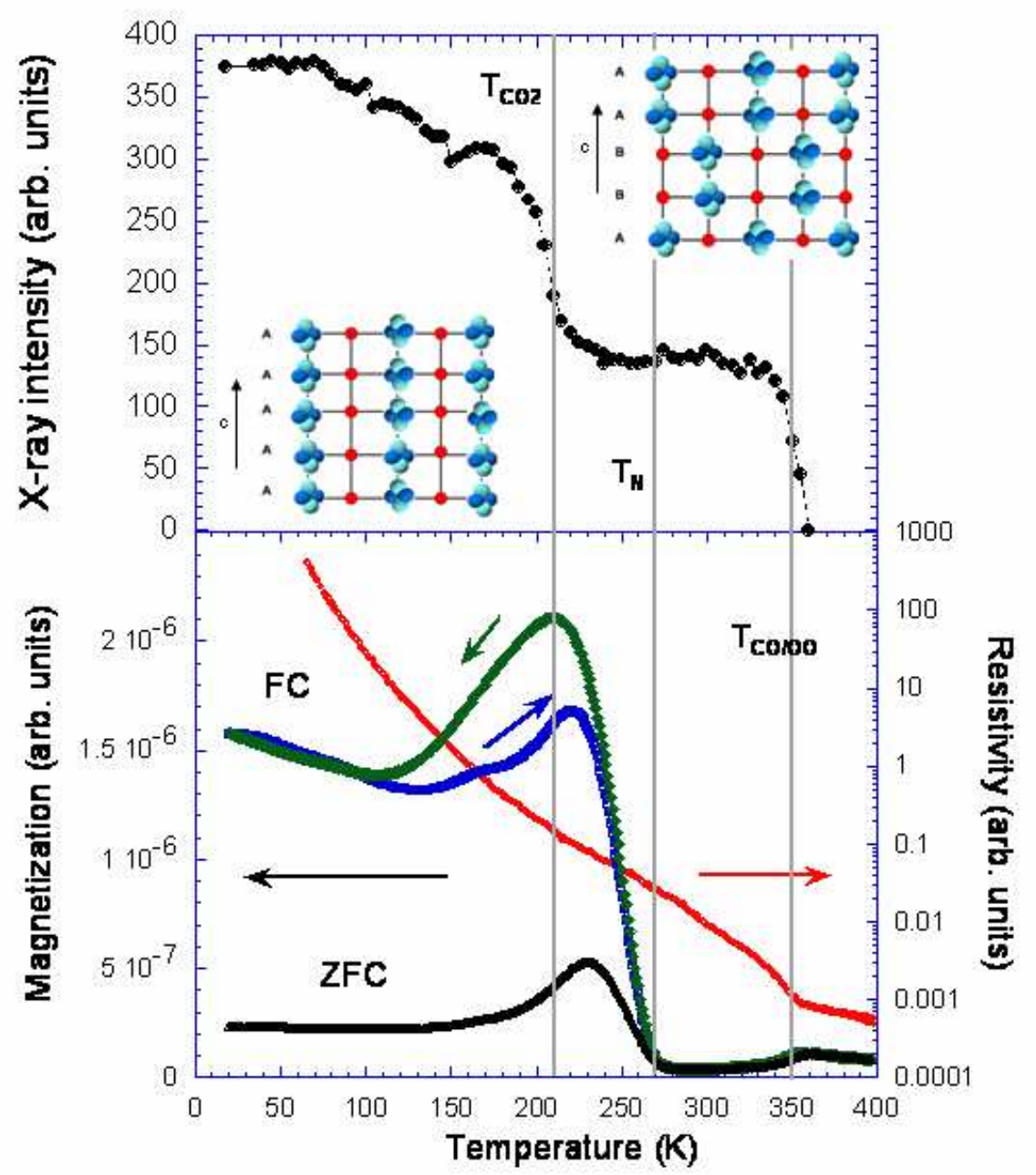

FIG. 3 (color online) a) Temperature dependence of the integrated x-ray intensity of the (¼ ¼ 0) reflection taken at the Mn $L_3$ edge (643.25 eV) with π incident radiation of polycrystalline $SmBaMn_2O_6$. The vertical lines indicate the various phase transitions as discussed in the text and the insets show the different orbital sackings b) Temperature dependence of the zero-field cooled (ZFC) and field cooled (FC) magnetization recorded in H=400A/m during cooling and heating (left axis) of polycrystalline $SmBaMn_2O_6$. In addition the temperature dependence of the resistivity is shown on a logarithmic scale (right axis).

We now discuss the structure factor of the (¼ ¼ 0) reflection for the various proposed stacking sequences along the c axis. The structure factor is described by a term with $(1+i)T_k^2$ for ABAB and AABB type stacking and by a term $2T_k^2$ for AAAA type stacking, where $T_k^2$ represents the second rank tensor for the $x^2$-$z^2$/$y^2$-$z^2$ type quadrupoles. As a result the intensity for the AAAA stacking is twice that of the ABAB/AABB type stacking, which is in excellent agreement with the observed increase in intensity at $T_{CO2}$. Therefore, our data support the AABB type stacking for T > $T_{CO2}$ as determined by electron diffraction by the observation of half integer *l* index reflections[6, 7], but indicate AAAA type stacking for T <

$T_{CO2}$, in contrast to the interpretation of the electron diffraction data. That the transition at $T_{CO2}$ is not associated with a change of orbital ground state can be appreciated from the unaltered energy dependency of the (¼ ¼ 0) reflection as shown in Figure 4. In particular no change is observed within the experimental accuracy between 180 and 235 K, which is below and above the orbital stacking transition respectively.

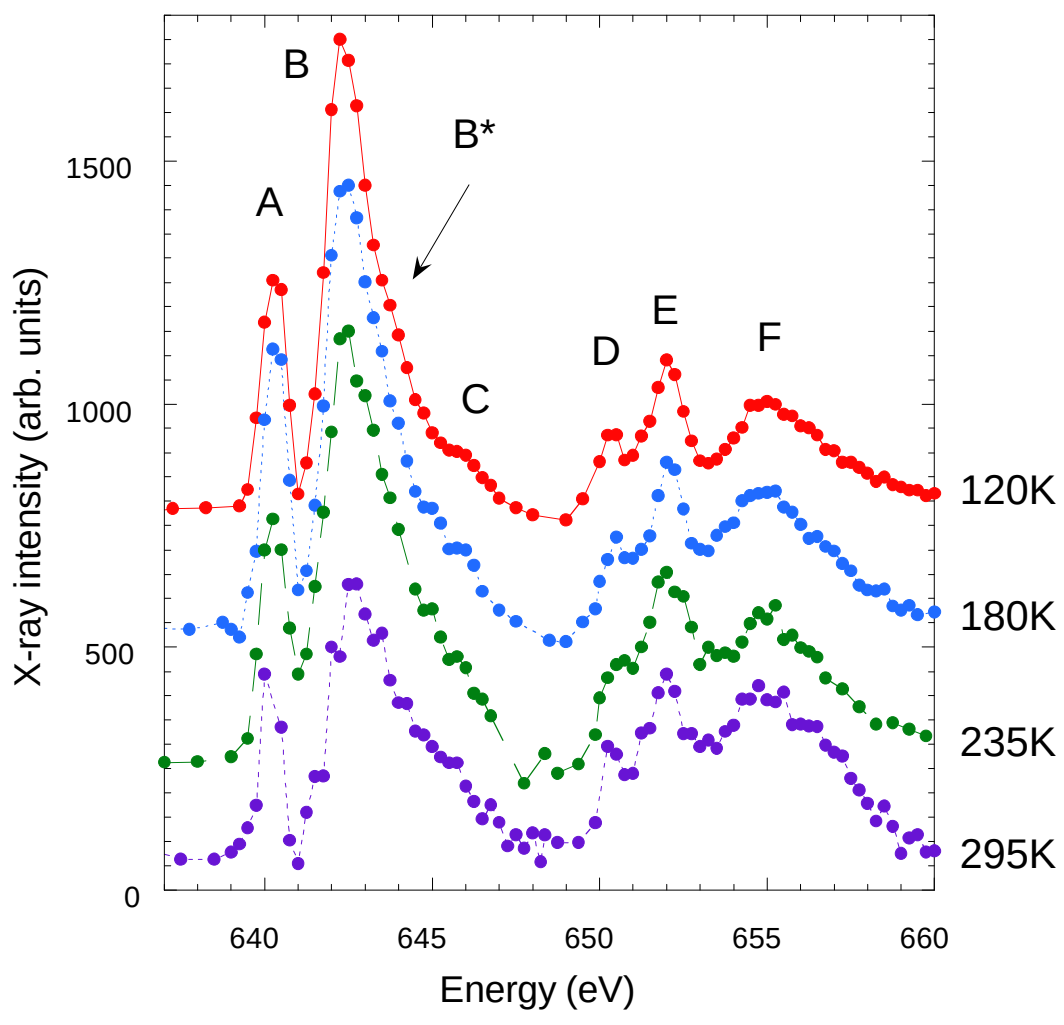

FIG. 4 (color online) Energy dependence of the integrated intensity of the (¼ ¼ 0) reflection of polycrystalline $SmBaMn_2O_6$ recorded at various temperatures with $\pi$ incident radiation. The intensities are rescaled and an offset is added to allow a better comparison between the spectra. The labeling of the energy features are the same as in Ref. [16].

For single-layer $La_{0.5}Sr_{1.5}MnO_4$, the temperature dependences of the various features in the resonant spectra are very distinct,[12, 16] in particular at the $L_2$ edge. The temperature dependence of feature F reflects the Jahn-Teller distortion while features D and E show a more regular order parameter due to the electronic order.[16] In strong contrast to these results, the energy dependence is constant for the layered $SmBaMn_2O_6$ in the whole temperature range and not affected by the reorientation transition nor the occurrence of antiferromagnetic order at T=250K. Therefore it is concluded that the Jahn-Teller distortion and the intrinsic electronic ordering are strongly coupled and show identical temperature dependence.

The integrated intensity of the (¼ ¼ 0) reflection shows a further increase below $T_{CO2}$, in contrast to its constant value between $T_{CO2}$ and $T_{CO}$. This increase of the charge and orbital ordering expectation values for decreasing temperatures in the AAAA ($T<T_{CO2}$) stacking phase indicates a further electron localization process for $T<T_{CO2}$, which is substantiated by the change in slope of the resistivity at this temperature. We infer that the ferromagnetic correlations due to double exchange interaction compete with the charge and orbital ordering above $T_{CO2}$ and inhibit a further growth of the order parameter. The change of stacking sequence at $T_{CO2}$ reduces the double exchange interaction as confirmed by the reduced magnetization, and leads to an increase of charge localization for $T<T_{CO2}$.

In conclusion, we present the first resonant soft X-ray powder diffraction experiment of an orbital order reflection to study the orbital ordering in A-site ordered layered manganite. The energy dependence of the (¼ ¼ 0) reflection indicates an $x^2$-$z^2$/$y^2$-$z^2$ type of orbital ordering due to the Sm/Ba layering and a strong correlation between the orbital order and the lattice. At $T_{CO2}$, a stacking reorientation transition to AAAA type of orbital stacking takes place which leads to a reduction of the double exchange and an increase of charge localization.

We thank the beamline staff of X11MA for their excellent support and the support of experimental equipment by J.M. Tonnerre and we thank K. Prsa for his support during the resistivity measurements. This work was supported by the Swiss National Science Foundation and the NCCR MaNEP Project and performed at SLS of the Paul Scherrer Insitut, Villigen PSI, Switzerland. We acknowledge financial support from the *Access to Major Research Facilities Programme* which is a component of the *International Science Linkages Programme* established under the Australian Government's innovation statement, *Backing Australia's Ability.*